\documentclass[twocolumn, superscriptaddress, prb, footinbib]{revtex4-1}

\usepackage{graphicx}
\usepackage{multirow}

\usepackage[version=3]{mhchem}
\usepackage{graphicx}
\usepackage{amsmath}
\usepackage{multirow}
\usepackage{color}

\begin{document}
\title{Understanding the Initial Stages of Reversible Mg Deposition
and Stripping in Inorganic Non-Aqueous Electrolytes}

\author{Pieremanuele Canepa} \email{pcanepa@mit.edu}
\affiliation{
Department of Materials Science and Engineering, Massachusetts
Institute of Technology, Cambridge, MA 02139, USA}

\author{Gopalakrishnan Sai Gautam}\affiliation{
Department of Materials Science and Engineering, Massachusetts
Institute of Technology, Cambridge, MA 02139, USA}

\author{Rahul Malik}\affiliation{
Department of Materials Science and Engineering, Massachusetts
Institute of Technology, Cambridge, MA 02139, USA}

\author{Saivenkataraman Jayaraman}\affiliation{
Department of Materials Science and Engineering, Massachusetts
Institute of Technology, Cambridge, MA 02139, USA}

\author{Ziqin Rong}\affiliation{
Department of Materials Science and Engineering, Massachusetts
Institute of Technology, Cambridge, MA 02139, USA}

\author{Kevin R. Zavadil}\affiliation{
Sandia National Laboratories, Albuquerque, NM 87185, USA}

\author{Kristin Persson}\affiliation{
Environmental Energy Technologies Division, 
Lawrence Berkeley National Laboratory, Berkley, CA 94720, USA}

\author{Gerbrand Ceder} \email{gceder@mit.edu}
\affiliation{
Department of Materials Science and Engineering, Massachusetts
Institute of Technology, Cambridge, MA 02139, USA}


\begin{abstract}
Multi-valent (MV) battery architectures based on pairing a Mg metal
anode with a high-voltage ($\sim$ 3 V) intercalation cathode offer a
realistic design pathway toward significantly surpassing the energy
storage performance of traditional Li-ion based batteries, but there are
currently only few electrolyte systems that support reversible Mg
deposition. Using both static first-principles calculations and \emph{ab
initio} molecular dynamics, we perform a comprehensive adsorption study
of several salt and solvent species at the interface of Mg metal with an
electrolyte of Mg$^{2+}$ and Cl$^-$ dissolved in liquid tetrahydrofuran
(THF). Our findings not only provide a picture of the stable species at
the interface, but also explain how this system can support reversible
Mg deposition and as such we provide insights in how to design other
electrolytes for Mg plating and stripping. The active depositing species
are identified to be (MgCl)$^+$ monomers coordinated by THF, which
exhibit preferential adsorption on Mg compared to possible passivating
species (such as THF solvent or neutral MgCl$_2$ complexes). Upon
deposition, the energy to desolvate these adsorbed complexes and
facilitate charge-transfer is shown to be small ($\sim$ 61 -- 46.2
kJ~mol$^{-1}$ to remove 3 THF from the strongest adsorbing complex), and
the stable orientations of the adsorbed but desolvated (MgCl)$^+$
complexes appear favorable for charge-transfer. Finally, observations of
Mg-Cl dissociation at the Mg surface at very low THF coordinations (0
and 1) suggest that deleterious Cl incorporation in the anode may occur
upon plating. In the stripping process, this is beneficial by further
facilitating the Mg removal reaction.
\end{abstract}

\maketitle

\section{Introduction}
\label{sec:intro}
Since their commercial introduction nearly twenty-five years ago,
rechargeable Li-ion batteries have performed admirably as the linchpin
technology enabling today's mobile electronics industry, currently
powering hundreds of millions of laptops, cameras, and phones
worldwide.\cite{Armand2008}  However, after years of continued
improvement in performance we are approaching a fundamental limit of
what can be accomplished with the current rocking chair Li-ion battery
technology platform.\cite{Thackeray2012} Further improvements in energy
density (the quantity of energy that can be stored per charge or
discharge either by mass or volume) without increasing the overall cell
cost are still required, not only to lengthen the per-charge battery
life of mobile devices, but also to facilitate the widespread adoption
of electrified vehicles and grid-scale renewable energy storage.  In
short, a disruptive innovation in electrochemical energy storage can
bring about tremendous technological and societal change--- it has the
realistic potential to displace significant CO$_2$ emissions from two of
the largest global industrial contributors in transport and electricity
generation.

One of the most promising approaches is to build on the traditional
robust three-component Li-ion battery architecture (intercalation or
metal anode, non-aqueous electrolyte, and intercalation cathode), but
with a multi-valent chemistry such as
Mg$^{2+}$,\cite{Aurbach2002,VanNoorden2014} where multiple electrons are
transported per ion (i.e.\ 2 electrons for each Mg$^{2+}$ compared to 1
for Li$^+$). This feature increases the energy density per intercalated
ion in an electrode, but equally important, along with the switch to a
Mg-based chemistry comes the potential to use a Mg metal anode as
opposed to an intercalation structure ($\sim$ 700 Ah/L for Li in
graphite compared to 3830 Ah/L for metallic Mg). For Li the use of
metallic anodes especially challenging due to safety
reasons.\cite{Cohen2000}

The path to fully functioning and cost-competitive Mg-ion batteries
begins with addressing a few but complex scientific questions, and
perhaps the most pressing is to develop a robust understanding of the
atomistic mechanism of reversible plating (or deposition) and stripping
(or dissolution) of Mg at the anode/electrolyte interface during battery
operation. To date, reversible Mg plating with low over-potential and
reasonable anodic stability has been achieved in practice with only a
specific class of electrolytes, namely organic or inorganic magnesium
aluminum chloride salts (magnesium-chloro complexes) dissolved in
ethereal solvents\cite{Gregory1990,Pour2011,Doe2014,Barile2014a,Barile2014,Yoo2013}.
Aurbach and collaborators developed the first operational Mg ion full
cells using electrolytes based on Grignard reagents (e.g.\
butyl-magnesium chloride),\cite{Gregory1990} and improved upon their
performance by switching to an in-situ prepared magnesium
organo-haloaluminate, which can achieve high coulombic efficiency
during stripping and deposition and anodic stability up to 3.0
V.\cite{Pour2011,Doe2014,Yoo2013} Recently, Muldoon and co-workers
further improved upon air-sensitivity and anodic stability by designing
a non-nucleophilic electrolyte containing a Hauser base, such as
hexamethyldisilazide magnesium chloride (HMDSMgCl) and
AlCl$_3$.\cite{Kim2011} Another functioning electrolyte, though
displaying limiting anodic stability, can be achieved using a
combination of Mg(BH)$_4$ and LiBH$_4$ in diglyme as demonstrated by
Shao \emph{et al.}\cite{Shao2013} Despite this substantial progress,
some aspects of the electrolytes are in need of improvement,
particularly the anodic stability, safety and compatibility with other
cell components.\cite{Doe2013} In contrast, most other non-aqueous
electrolytes, including the Mg analogues to the common Li-ion battery
electrolytes, lead to immediate electrolyte decomposition on the Mg
anode, which passivates the metal surface preventing further
electrochemical reactions.\cite{Aurbach2002}

Much effort has been devoted toward characterizing the precise sequence
of atomic-scale processes that make up the overall Mg
deposition/dissolution process in ethereal solvents.  First, extensive
experimental and computational approaches have been used to characterize
the equilibrium solvation structure of Mg$^{2+}$ in the bulk
electrolyte, which defines the thermodynamic start (end) point of
deposition (dissolution), and several possible coordinating complexes
involving combinations of Mg$^{2+}$, Cl$^-$, and THF have been
identified.\cite{Pour2011,Doe2014,Barile2014,Yoo2013,Gizbar2004,Wan2014,Benzmayza2013,Muldoon2014}
These findings have subsequently been used to inform phenomenological
models that describe the general sequence of ion desolvation, adsorption
on the anode surface, charge-transfer, and metal incorporation that make
up the overall deposition process (the sequence in reverse describes the
dissolution process) see Ref.~\citenum{Yoo2013} and references therein.
Only if all of these individual processes are kinetically unobstructed
an Mg anode can operate reversibly and efficiently, e.g.\ with low
overpotential and high coloumbic efficiency. 

However, the complete understanding at the atomic scale of the processes
of electrochemical plating and stripping at the Mg anode/electrolyte
interface, which would enable rational design of the next-generation
Mg-based electrolytes, is still lacking.  In recent work, Doe \emph{et
al.}\cite{Doe2014} showed that purely inorganic-based electrolytes
solutions (i.e.\ without organometallic moieties), also referred as the
``MACC'' electrolyte, can exhibit reversible Mg electro-deposition with
anodic stability up to 3.1 V \emph{vs.}\ Mg.  Hence, the MACC electrolyte
system can serve as an excellent prototype system to understand Mg
deposition and stripping in the presence of Mg-Cl complexes. Here, we
present an in-depth first-principles study aimed at characterizing the
relevant chemical structures and complexes that form at the interface of
a Mg metal anode in contact with an electrolyte solution comprised of
Mg$^{2+}$ and Cl$^-$ dissolved in an ether-based solvent
(tetrahydrofuran or THF). By performing static first-principles
adsorption calculations combined with \emph{ab initio} molecular
dynamics (AIMD) simulations, we are able to separately probe the
interaction of the THF solvent in contact with the anode surface, as
well as the full electrolyte/anode interaction by incorporating both
Mg$^{2+}$ and Cl$^-$.  Combined, these calculations reveal important
features about magnesium-chloro complexes and their ability to function
in practice:  first, cyclic ether-based solvents are confirmed to be
chemically inert against the Mg anode, while polymer-like
ethers display comparatively more passivating behavior on the Mg anode;
and second, our calculations reveal the existence of several (MgCl)$^+$
coordination complexes similar in energy but solvated by different
numbers of THF molecules that define a kinetically facile Mg$^{2+}$
desolvation process during deposition.  

The computational and theoretical strategy used to gain insights
into the species at the Mg anode electrode and Mg
desolvation/dissolution is non-specific and can be readily applied to a
variety of important problems such as electroforming of metals,
corrosion of metals and alloys, medicinal chemistry, and catalysis. We
emphasize that to obtain physically meaningful results, we consider
explicitly the interaction of non-aqueous liquids with a solid phase,
which is commonly approximated by a vapor/solid interface in other
studies.


\section{Methodology}
\label{sec:method}
Absorption calculations and \emph{ab initio} molecular dynamics (AIMD)
simulations were performed within the DFT approximation, using the
vdW-DF functional\cite{Langreth09,Thonhauser07} as implemented in
VASP.\cite{Kresse93,Kresse96} The total energy was sampled on a
well-converged 3$\times$3$\times$1 \emph{k}-point grid together with
projector augmented-wave theory\cite{Kresse99} and a 520 eV plane-wave
cutoff, and forces on atoms were converged within 1$\times$10$^{-2}$ eV
\AA$^{-1}$. For the different adsorption geometries of
Figure~\ref{fig:adsorption} and Figure~\ref{fig:binding}a the adsorption
energy ($\Delta$E) is defined as: 
\begin{equation}
\label{eq:deltae}
\Delta E = E_{\rm Surf.+Ads.}{\rm(s)} - E_{\rm Surf.}{\rm(s)} -E_{\rm Ads.}{\rm(l)}\;,
\end{equation}
where the $E_{\rm Surf.+Ads.}{\rm(s)}$ is the total energy of the
species formed by the Mg-surface and the adsorbing molecule (solvent or
salt) in the solid state, $E_{\rm Surf.}{\rm(s)}$ the total energy of
the surface in the solid state, and $E_{\rm Ads.}{\rm(l)}$ the total
energy of the $n$ adsorbing species.  The $E_{\rm Ads.}{\rm(l)}$  is
conventionally  approximated by the total energy of its  gas reference,
but to address the specific problem of the electrode in contact with a
liquid electrolyte it is more practical to refer to the liquid state.
To compute the liquid reference $E_{\rm Ads.}{\rm(l)}$  we proceed by
isolating low-energy snapshots from AIMD runs of both salt (MgCl$^+$ or
MgCl$_2$ in THF solvent) and solvent (THF) and optimize their
geometries. The average total energy provides a good estimate of the
liquid reference and incorporates both stabilizing 1st solvation shell
effects and vdW-effects from the solvent. 

As the coordination of (MgCl)$^+$ by THF in bulk is still debated, two
different liquid (MgCl)$^+$ bulk references were computed: \emph{i})
four coordinated Mg, (MgCl)$^{+}$-3THF as determined by Wan \emph{et
al.}\cite{Wan2014} and \emph{ii}) six-coordinated Mg, (MgCl)$^{+}$-5THF
as proposed by previous experimental
work.\cite{Doe2014,Barile2014,Yoo2013,Gizbar2004,Benzmayza2013}

For the AIMD calculations the cutoff was reduced to 400 eV and the total
energy was only sampled at the $\Gamma$-point. Using a sampling of 1 fs,
a production period of 15 ps was preceded by an equilibration period of
20 ps within the canonical ensemble (NVT) at 300 K, using the
Nos\'{e}-Hoover thermostat.\cite{Nose1984,Hoover1985} The slabs for the
adsorption calculations were cleaved from the fully relaxed geometry of
the Mg bulk (with \emph{a} = 3.202 \AA\ and \emph{c} = 5.139~\AA)
and MgO (with \emph{a} = 4.201 \AA). Mg (0001) and MgO (100)
surfaces were modeled using a 5 layer periodic slab. Specifically for
the adsorption of THF and poly-THF a 4$\times$4 (\emph{a} = \emph{b} =
12.808 \AA) slab with 80 Mg atoms was employed, whereas  a
2$\times$2  (\emph{a} = \emph{b} = 8.931 \AA) slab with 90 atoms for
MgO. For the adsorption of (MgCl)$^{+}$ and MgCl$_2$ complexes, a
larger surface with size 5$\times$5 (\emph{a} = \emph{b} = 16.010 \AA)
and 125 Mg atoms is needed. A well-converged vacuum of 35 \AA\ was
inserted between surface images, and the adsorbing species were always
adsorbed on both surfaces to minimize artificial electric dipoles
through the non-periodic direction of the slab. The atomic positions of
the two most exposed layers of the slabs and the adsorbates were
optimized. All the geometries employed in the AIMD simulations were
created by inserting a number of THF molecules at the experimental
density (0.889~g/cm$^3$) in the vacuum space between slabs to simulate
liquid, resulting in unit cells with  $\sim$ 550 atoms. To accelerate
computationally costly AIMD simulations, we performed an intermediate 1
ns pre-equilibration NVT classical molecular dynamics simulation using
the CHARMM potential as implemented in
DL\_POLY\cite{MacKerell1998,Todorov2006} while keeping the coordinates
of the surface and adsorbate atoms fixed as obtained from preliminary
adsorption calculations. The concentration of the MgCl$^+$ salt in the
AIMD is 0.30 M and falls in the range of the experimental concentrations
0.25 -- 0.40 M explored by Aurbach and collaborators.\cite{Mizrahi2008}
Diffusivities were extracted from the mean squared displacements of the
molecules at the surface from the AIMD simulations.

\section{Results}
\label{sec:results}
\subsection{THF on Mg}
\label{subsec:THF}
\begin{figure}[t]
\includegraphics[width=\columnwidth]{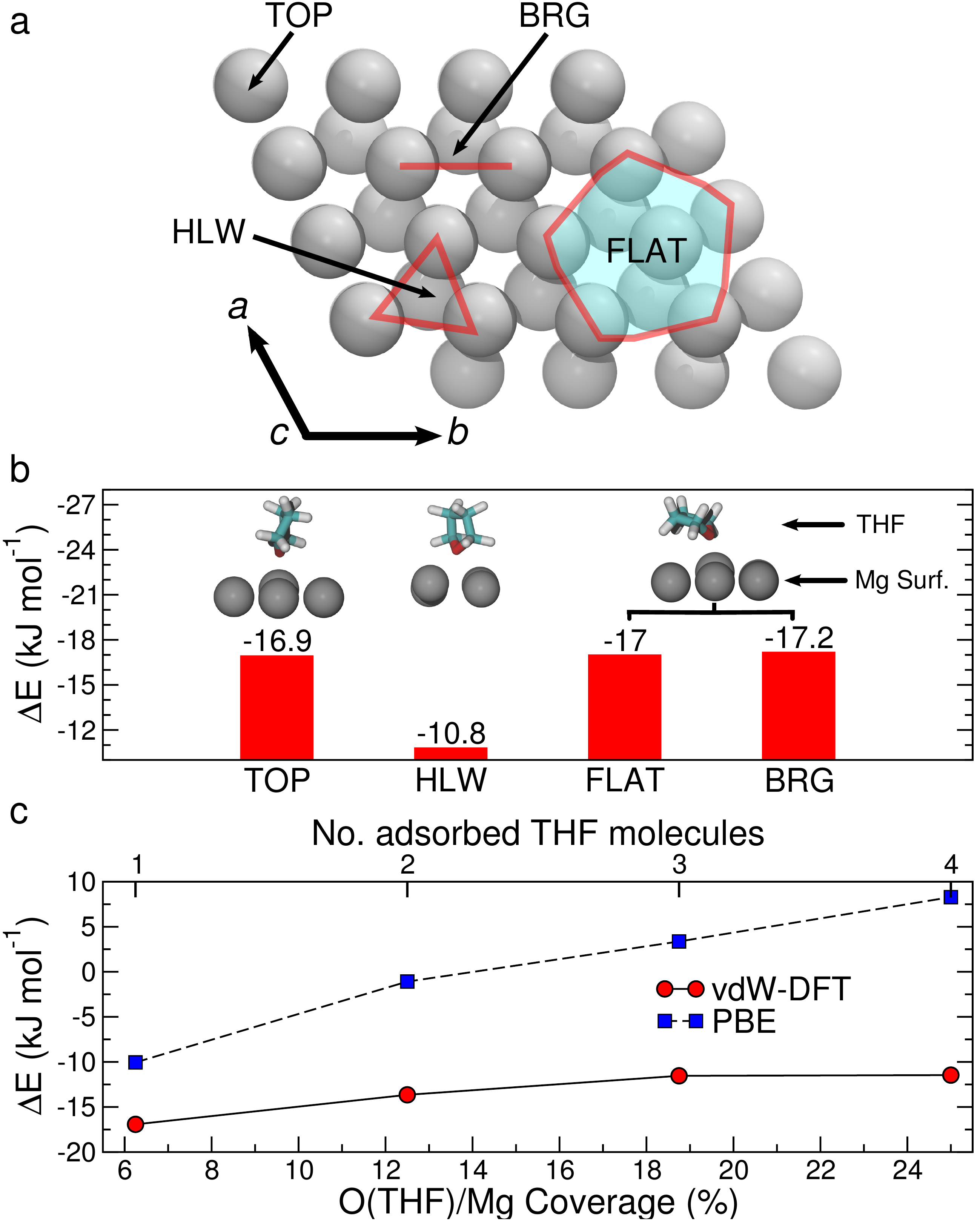}
\caption{\label{fig:adsorption} (Color online) (a) Stable adsorption
sites (TOP, HLW, FLAT, and BRG) for THF on Mg (0001), and corresponding
(b) adsorption energies $\Delta$E (in kJ~mol$^{-1}$ referenced to the
THF liquid state) with snapshots of their fully relaxed THF
configurations (inset). (c) Variation of the adsorption energy with THF
surface coverage O(THF)/Mg.}
\end{figure}

Adsorption energies computed from first-principles calculations are used
to quantify the chemical interaction of a solvent tetrahydrofuran (THF)
molecule with the Mg metal anode. From the standpoint of electrostatic
interaction, the THF molecule is expected to preferentially align with
the ethereal oxygen oriented toward the exposed Mg sites at the surface.
Previous experimental{\cite{Matsui2011}} and
theoretical{\cite{Ling2012}} work has demonstrated that Mg deposition
and growth occur preferentially on the Mg (0001) facet at low current
densities ($\sim$ 0.5 $-$ 1.0 mA~cm$^{-2}$), hence the analysis is
restricted to the Mg (0001) surface. Figure~\ref{fig:adsorption}a
and~\ref{fig:adsorption}b show four plausible adsorption sites at the Mg
(0001) surface for a THF molecule labeled TOP, BRG, HLW, and FLAT, and
their adsorption energies ($\Delta$E). In this manuscript we refer
strictly to the hcp hollow site HLW as indicated by the triangle in
Fig.~\ref{fig:adsorption}a as opposed to the less stable fcc hollow
site, which shows to have consistently weaker adsorption energies.

The calculated adsorption energies of Figure~\ref{fig:adsorption}b
(referenced to liquid THF) fall into a narrow range between -10 and -17
kJ~mol$^{-1}$ with negligible variation between distinct surface sites
and are consistent with the inert nature of ethereal molecules in the
presence of a highly electropositive
metal.\cite{Aurbach2002,Aurbach2011,Yoo2013} Negative $\Delta$Es
indicate that spontaneous adsorption is preferred.\break Referencing to
the THF gas state would make the adsorption energies more negative (see
Supplementary Information Figure\ S1) by adding the gas to liquid
bonding energy ($\sim$~-46~kJ~mol$^{-1}$). The lower $\Delta$E
($\sim$~-17.2 kJ~mol$^{-1}$) of the FLAT configuration originates from
the THF molecule's orientation at the surface, favoring not only the
O-Mg(surface) interaction (of the top configuration) but also additional
weaker C-Mg(surface) and H-Mg(surface) bonds. THF molecules initially
adsorbed in the BRIDGE sites are found to relax to the FLAT
configuration (see Figure~\ref{fig:adsorption}b). 

A more realistic picture of THF adsorption, consistent with experimental
conditions, is achieved by considering the effect of THF coverage.
Figure~\ref{fig:adsorption}c shows the variation of the adsorption
energies for increasing number of adsorbed THF. The interaction of THF
with the surface noticeably decreases with increasing THF coverage, and
the adsorption energies shift to further positive values consistent with
the occurrence of THF-THF lateral electrostatic repulsions. As a
comparison, we also report $\Delta$Es that do not include van der Waals
interactions (marked PBE in see Figure~\ref{fig:adsorption}c), which
provide a helpful upper-bound limit for the adsorption energies. The
entropic contributions originating from the THF adsorption from the
liquid will further increase the adsorption free energies to more
positive values, displacing the THF molecule from the Mg-anode and
ensuring that ether molecules do not passivate the surface. 

\begin{figure}[t]
\includegraphics[width=\columnwidth]{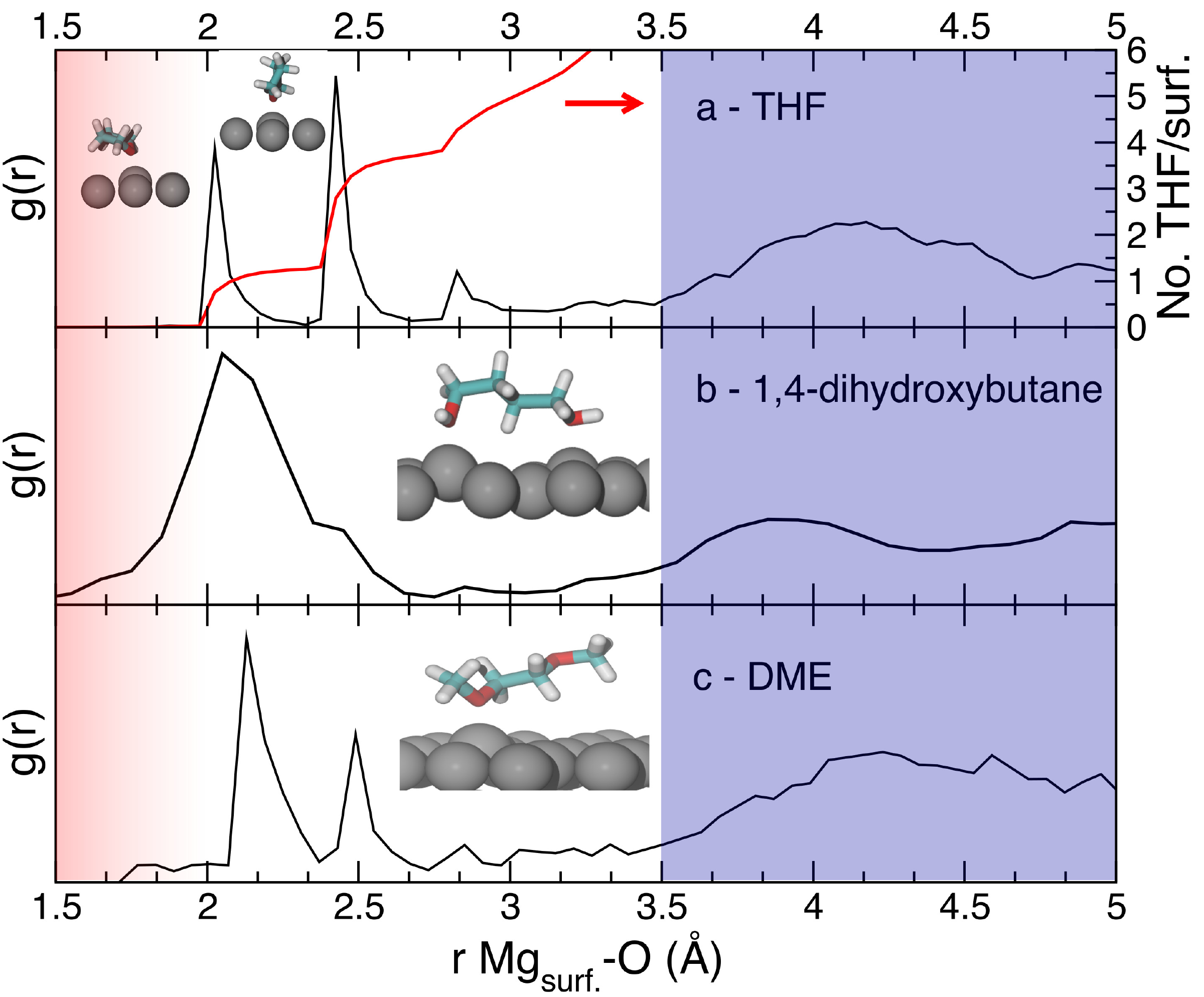}
\caption{\label{fig:THFRDF} (Color online) PDF, g(r) of ethereal oxygen
of THF and exposed Mg (0001) surface atoms (black lines), and surface
coverage (dashed red line) for (a) THF, (b) 1,4-dihydroxybutane, and
(c) DME (1,2-dimethoxyethane)} from AIMD simulations. Insets are
representative snapshots of THF and poly-THF fragments at the Mg
surface. Red and blue shades highlight the lower bound of PDFs and the
2$^{nd}$ solvation shell, respectively.
\end{figure}

To test the findings provided by the adsorption calculations and further
resolve the dynamic environment of the anode/solvent interface, AIMD
simulations within the canonical ensemble were performed at room
temperature (300 K) and at the experimental density of THF (0.889
g/cm$^{3}$).\cite{Mizrahi2008}  Figure~\ref{fig:THFRDF}a (black line)
shows the pair distribution function, PDF or g(r), of the exposed Mg
surface sites versus the THF oxygen atoms, Mg$_{\rm surf.}$-O. 

The two dominating peaks observed at $\sim$ 2 and 2.5~\AA\ show the
co-existence of two THF configurations at the Mg(0001) surface, FLAT
(peak at $\sim$ 2.0~\AA) and TOP (peak at $\sim$ 2.5~\AA), which also
have the most negative adsorption energies, as seen in
Figure~\ref{fig:adsorption}b. By integrating the PDF profiles of
Figure~\ref{fig:THFRDF}a (red line) we conclude that a maximum of four
THF molecules (2 in FLAT and 2 in TOP conformation) per 16 Mg exposed
surface sites can be simultaneously adsorbed (dashed red line in
Figure~\ref{fig:THFRDF}a), consistent with the adsorption results of
Figure~\ref{fig:adsorption}c. The diffusivity of THF molecules at the
surface extracted from their mean squared displacement observed in the
AIMD simulation is (6.1 $\pm$ 1.5) $\times$ 10$^{-6}$ cm$^2$~s$^{-1}$.
Physically, the mean squared displacement of a molecule at the metal
surface is related to the strength of adsorption--- if the interaction
is very strong (attractive), molecules near the surface remains
predominantly in the adsorbed state rather than in the liquid, resulting
in lower surface diffusivity.  


\subsection{poly-THF on Mg}
\label{subsec:polyTHF}
It is well understood that either electrophilic attack by protons on the
ethereal carbon (self-polymerization) or catalytic interaction with an
electropositive metal (such as Al in AlCl$_3$ or Mg) can initiate the
THF ring opening, followed by polymerization forming
poly-THF.\cite{Barile2014,Dreyfuss1967,Chisholm1996} We therefore also
investigate the possible adsorption of THF products of polymerization
such as poly-tetrahydrofuran (poly-THF) at the anode/electrolyte
interface shown in the inset of Figure~\ref{fig:THFRDF}b. 
Here, two approximations are used to
describe poly-THF: \emph{i})  by cleaving THF at the ethereal bonds and
saturating by hydrogen atoms, forming the 1,4-dihydroxybutane (see
Figure~{\ref{fig:THFRDF}}b), and \emph{ii}) by 1,2-dimethoxyethane also
known as glyme (DME, see Figure~{\ref{fig:THFRDF}}c).  While the acidic nature of the terminal
hydrogens in 1,4-dihydroxybutane may enhance the adsorption energy with
the Mg-surface compared to linear ethers (e.g.{\ } glyme,
tetraglyme, $n$glyme), it nonetheless provides an upper bound to the
adsorption energies of a variety of THF polymerization products that
might be present in the cell. The adsorption energies of
1,4-dihydroxybutane (referencing to a dilute solution of 0.4 M
1,4-dihydroxybutane-THF in liquid THF) for different configurations were
always found to be more negative (-32.0~$\pm$~5~kJ~mol$^{-1}$) than
those of cyclic THF.  Examination of the Mg$_{\rm surf.}$-O  PDF reveals
that the oxygen atoms of 1,4-dihydroxybutane are
always nearer to the Mg-surface (broad peak from 1.7 to 2.7 \AA) in
comparison to cyclic THF molecules (see Figure~\ref{fig:THFRDF}a, b),
likely due to more geometric degrees of freedom accessible by
1,4-dihydroxybutane that can allow for improved
binding for a linear molecule on the anode surface.  Measurements of the
diffusion coefficients (from AIMD simulations) further indicate a
smaller mobility of 1,4-dihydroxybutane at the
surface, $\sim$ (9.5 $\pm$ 1.6) $\times$10$^{-7}$ cm$^2$ s$^{-1}$, as
compared to weakly physisorbed and more mobile THF molecules, $\sim$
(6.1 $\pm$ 1.5) $\times$ 10$^{-6}$ cm$^{2}$ s$^{-1}$. However, at
the Mg(0001) surface DME is found to bind less strongly
(-10.5~$\pm$~1~kJ~mol$^{-1}$) than 1,4-dihydroxybutane
(-32.0~$\pm$~5~kJ~mol$^{-1}$), and comparable to THF
(-17.2~kJ~mol$^{-1}$), confirming the inert nature of this molecule.
Although the similarity of the Mg$_{\rm surf.}$-O PDFs of DME and THF
(see Fig.~{\ref{fig:THFRDF}}c and Fig.~{\ref{fig:THFRDF}}a) attests
comparable binding nature for these molecules, the minimum distance
between the ethereal oxygen of DME and Mg atoms at the surface falls at
slightly larger distances (see peak at 2.1 \AA{}) than in THF (see peak
at 2.0~\AA{} in Fig.~{\ref{fig:THFRDF}}a), which can be explained by the
larger steric hindrance of DME.

Recently, Barile \emph{et al.}\cite{Barile2014} have observed that the
addition of poly-THF in the MACC electrolyte increases the deposition
overpotentials to 500 mV indicating that poly-THF may passivate more the
Mg-surface.  We conclude that hydroxyl-terminated polymer molecules
(e.g.{\ } 1,4-dihydroxybutane)  are more prone to passivate the Mg-anode surface
(than linear-ethers and THF), thereby negatively affecting the battery
performances during stripping and deposition. Furthermore, the
chelating nature of linear ether with long chains (e.g.{\ } glyme, diglyme,
tetraglyme, and poly-THF) is expected to inhibit Mg$^{2+}$ ion delivery
at the surface.{\cite{Okoshi2013,Rajput2015}}


\subsection{THF on MgO}
\label{subsec:MgO}
The irreversible reaction of Mg with ubiquitous water forms
resistant passivation layers that suppress all Mg electrochemical
activity and is one of the limiting factors in the development of
Mg-ion batteries. Investigations of Mg corrosion have clarified the
bilayer structure of the passivating film, with the inner part
comprised of MgO and an external layer of
Mg(OH)$_2$.{\cite{Gofer2003,Song2011,Aurbach2011}} In order to clarify
the effect of surface passivation on solvent adsorption, we calculated
the adsorption energies for THF on the MgO(100)
termination,{\cite{Namba1981}} which is found to grow epitaxially on the
Mg(0001) facet. Analogous to previous sections we use static DFT
adsorption calculations to address the interaction of THF with MgO(100),
and verify these findings with auxiliary \emph{ab initio} MD
simulations.   We anticipate that the ethereal oxygen of THF will
interact preferentially with Mg atoms at the surface (as also seen by
THF on Mg metal) while the THF's H atoms will maximize H-bonding with
surface oxygens. 

\begin{figure}[t]
\includegraphics[width=\columnwidth]{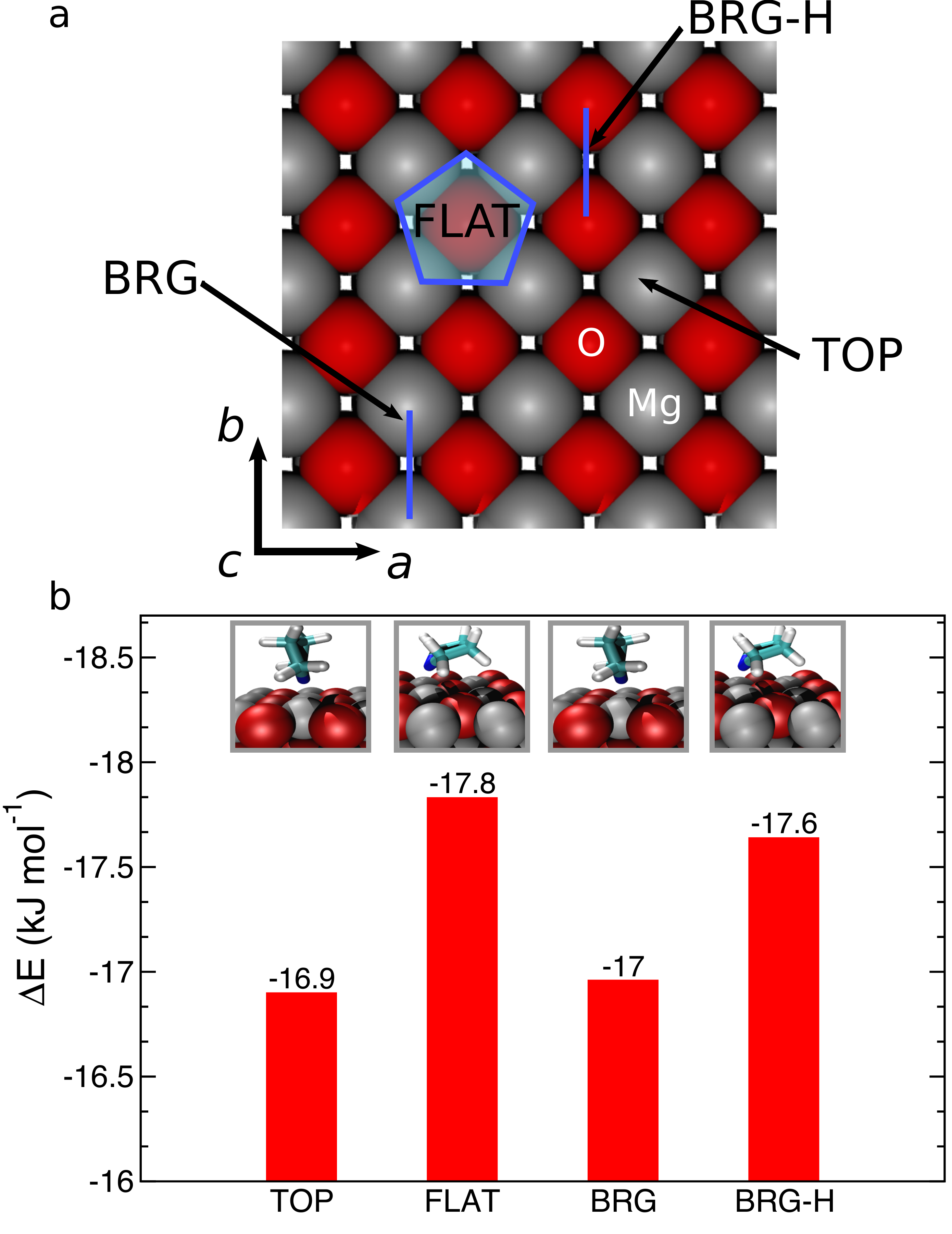} 
\caption{\label{fig:MgOall} (Color online) (a) Stable adsorption sites
(TOP, FLAT, BRG and BRG-H) for THF on MgO(100), and corresponding (b)
adsorption energies $\Delta$E (in kJ~mol$^{-1}$ referenced to the THF
liquid state) with snapshots of their fully relaxed THF configurations
(inset). THF ethereal oxygen in blue for clarity. }
\end{figure}

Figure~{\ref{fig:MgOall}}a and~{\ref{fig:MgOall}}b show four plausible 
adsorption sites at the MgO(100) surface for a THF molecule labeled TOP,
FLAT, BRG and BRG-H and their respective adsorption energies
($\Delta$E). BRG-H represents a THF adsorption arrangement that maximize
H-bonding between THF hydrogen atoms and MgO oxygen species at the
interface (as seen in Fig.~{\ref{fig:MgOall}}a). TOP, FLAT, and BRG
configurations are the same as described for adsorption on Mg. 

Surprisingly, the adsorption energies reported in
Fig.~{\ref{fig:MgOall}b} (referenced to liquid THF) are found remarkably
close to $\Delta$Es computed for THF on Mg(0001), varying between -17
and -18 kJ~mol$^{-1}$. By close examination of each THF adsorption
configuration and relative $\Delta$Es, we observe that the BRG
configuration relaxes to TOP, and similarly the FLAT configuration
relaxes to BRG-H, which maximizes the interactions with the surface
oxygen atoms through hydrogen bonds (see snapshots of
Fig.~{\ref{fig:MgOall}}b).  On the basis of these results we cannot
observe a substantial difference between the adsorption
properties of THF on MgO (100) and Mg (0001) surfaces and suggest that
further theoretical and experimental investigations are needed.

\begin{figure}[t]
\includegraphics[width=\columnwidth]{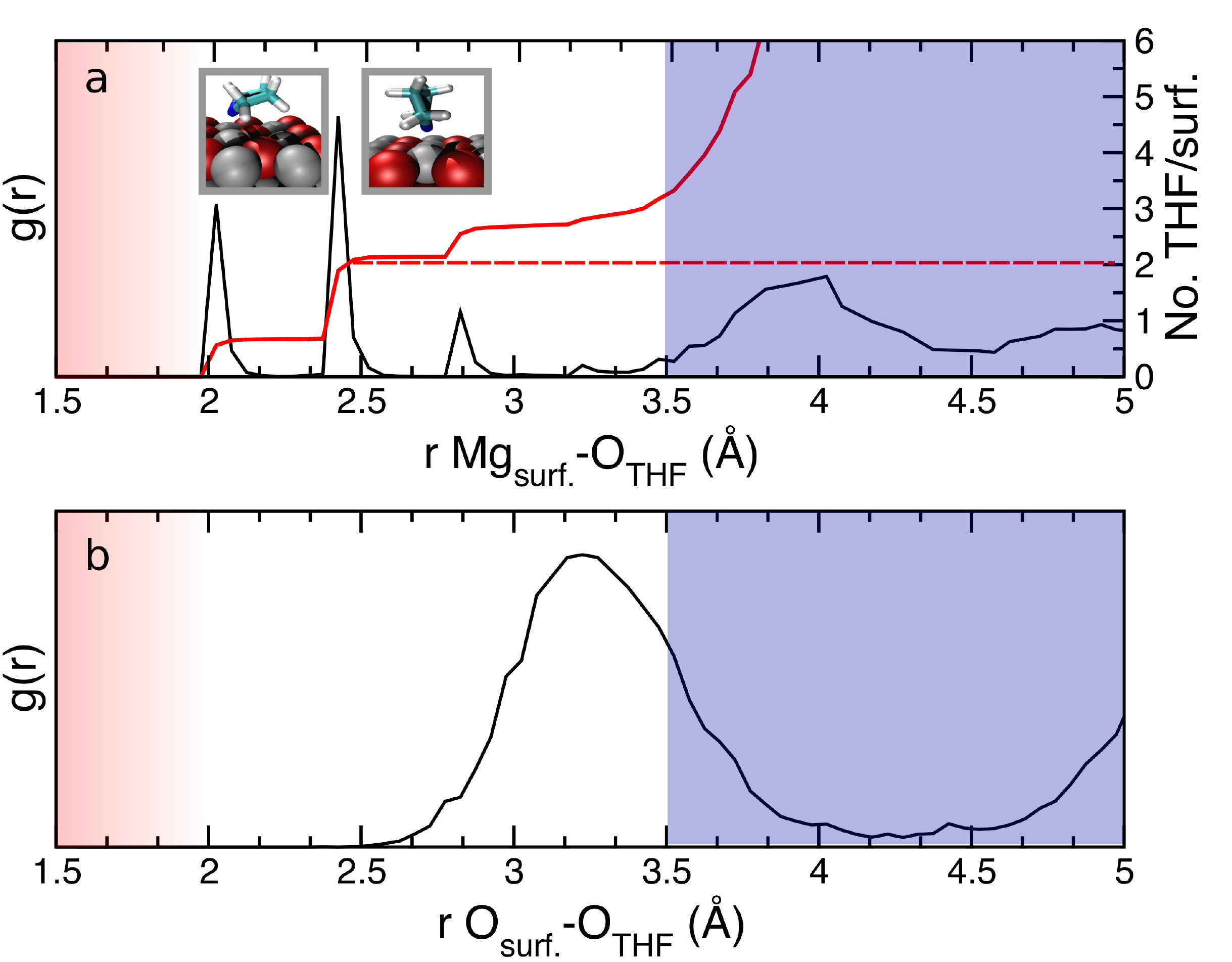} 
\caption{\label{fig:MgOPDF} (Color online) (a) PDF, g(r) of ethereal oxygen
of THF and exposed Mg surface atoms and (b) O exposed surface atoms on MgO(100) (black lines), respectively. Coverage (dashed red line). Insets are representative snapshots of THF fragments at the MgO surface. Red and blue shades highlight the lower
bound of PDFs and the 2$^{nd}$ solvation shell, respectively }
\end{figure}

AIMD simulations are performed to investigate the dynamic properties
of THF molecules at the MgO interface with similar settings described in
previous sections. Figure~{\ref{fig:MgOPDF}}a and~{\ref{fig:MgOPDF}}b
(black line) show the pair distribution functions, PDF or g(r), of the
exposed Mg and O surface sites versus the THF oxygen atoms, Mg$_{\rm
surf.}$-O$_{\rm THF}$ and O$_{\rm surf.}$-O$_{\rm THF}$, respectively. 

Similar to Fig.~{\ref{fig:THFRDF}}, the Mg$_{\rm surf.}$-O$_{\rm
THF}$ PDF of Fig.~{\ref{fig:MgOPDF}}a shows two major peaks at $\sim$ 2
and 2.5 {\AA}, emphasizing the co-existence of two THF adsorption
configurations at the MgO(100) surface: BRG-H (peak at $\sim$ 2.0 \AA)
and TOP (peak at $\sim$ 2.5 \AA). The remarkable similarity of the
Mg$_{\rm surf.}$-O PDFs on Mg(0001) and MgO(100) (seen comparing
Fig.~{\ref{fig:THFRDF}}a and Fig.~{\ref{fig:MgOPDF}}a) suggests that PDF
analysis is not sufficient to distinguish whether THF molecules are
adsorbed on active or passivated Mg anodes. The PDF O$_{\rm
surf.}$-O$_{\rm THF}$ shown in Fig.~{\ref{fig:MgOPDF}}b is more
informative and can discriminate between adsorption on active or
passivated surfaces, with a single peak between 2.7 and 4 {\AA{}}
characteristic to absorption on MgO.  Mg$_{\rm surf.}$-O$_{\rm THF}$ PDF
also provides information about the THF coverage on MgO (see red line in
Fig.~{\ref{fig:MgOPDF}a}),  concluding that a maximum of two THF
molecules (1 in FLAT and 1 in BRG-H conformation) can be simultaneously
adsorbed  per 12 Mg exposed surface sites (dashed red line in
Fig.~{\ref{fig:MgOPDF}}a).


\subsection{Magnesium-chloro Salt and THF solvent}
\label{subsec:salt}
By first establishing that ethereal solvent molecules do not
significantly interact with the Mg anode surface, we turn our attention
to characterizing the anode/electrolyte interface by including the
magnesium chloride salt species (solvated by THF).  The atomic-scale
interactions between salt and solvent that define the relevant reacting
complexes and species in the bulk electrolyte can be quite complex as
demonstrated in seminal work by Aurbach and collaborators, who
characterized intricate equilibria between magnesium-chloro clusters and
AlCl$^{4-}$ in the form of ``simpler'' monomers
[Mg($\mu$-Cl)$\cdot$5THF]$^{+}$[(AlCl$_4$)]$^{-}$, dimers
[Mg$_2$($\mu$-Cl)$_3\cdot$6THF]$^+$[(AlCl$_4$)]$^-$, and larger
polymeric
species.\cite{Pour2011,Doe2014,Barile2014,Gizbar2004,Benzmayza2013} A
recent theoretical work by Wan \emph{et al.}\cite{Wan2014} clarified the
first solvation shell of the monomer and dimer magnesium organo-chloro
species in the bulk electrolyte. Their AIMD calculations suggest that
the (MgCl)$^+$ monomer is always coordinated by three THFs. However, it
is still debated whether the (MgCl)$^+$ monomer is actively present in
the bulk electrolyte or only formed in proximity of the electrode as the
product of the dimer decomposition.\cite{Benzmayza2013} The structure of
the organo-chloroaluminate salts
(C$_2$H$_5$MgCl-[(C$_2$H$_5$)$_2$AlCl]$_2$ in THF, DCC) in
proximity of the Mg-anode was investigated experimentally using XANES
spectroscopy by Benzmayza \emph{et al.}\cite{Benzmayza2013} revealing
that only the cation monomer [Mg($\mu$-Cl)$\cdot$5THF]$^+$ can approach
the surface (as opposed to the more complex and bulkier dimer and
trimer), thus representing the active species during the Mg-deposition.
Previous work{\cite{Gizbar2004,Pour2011,Benzmayza2013,Barile2014}
on the organo and inorgano chloroaluminate electrolytes have elucidated
the complex thermodynamic equilibria between monomer, dimer and larger
multimeric species, which might ultimately affect the concentrations of
the (MgCl)$^+$ near the anode electrode.} Even though more complex
Mg-carriers (e.g. dimer and larger multimeric units) could also be
present at the surface, we set to investigate only the monomer
(MgCl)$^+$ as the simplest possible vehicle to transport Mg$^{2+}$ ions
from the electrolyte in proximity of the anode.{\cite{Barile2014}} In
this work, we use these findings to inform the construction and
preparation of our first-principles simulations, focusing on the
interaction of (MgCl)$^+$ monomers coordinated by THF near the anode
surface. 

\begin{figure*}[t]
\includegraphics[scale=0.25]{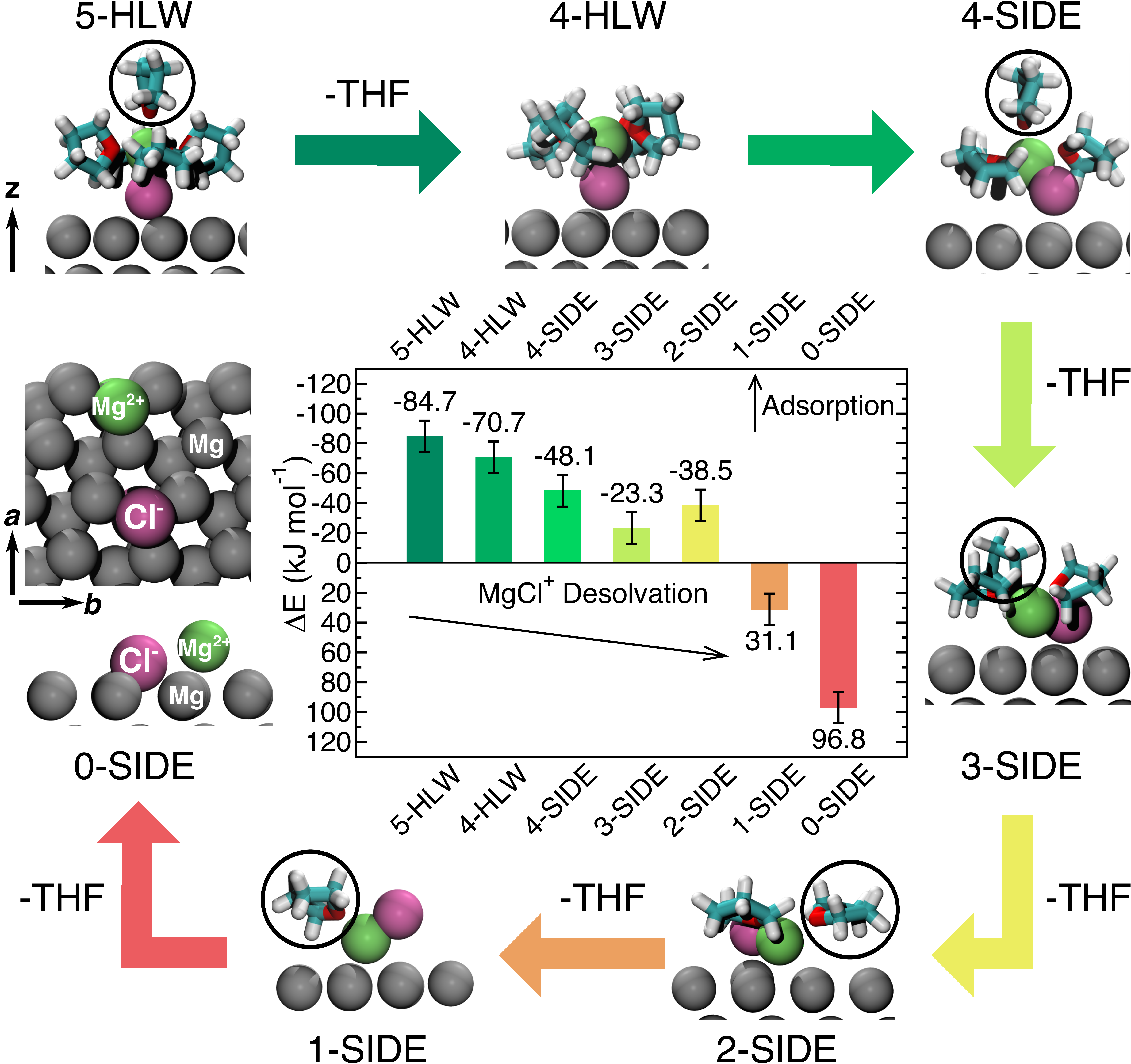}
\caption{\label{fig:binding} (Color online) stable configurations of
(MgCl)$^+$ solvated by varying No. of THF (from 5 to 0) adsorbed at the
Mg (0001) surface, and corresponding adsorption energies ($\Delta$E in
kJ~mol$^{-1}$). Upper and lower bounds indicated by the bars on
$\Delta$Es are from using 3-THF and 5-THF bulk liquid reference states,
respectively. Black circles and color-coded arrows indicate THF
removal.}
\end{figure*}

Using a similar approach to the one used to study the anode/solvent
interface, we first probe the adsorption energies of possible
magnesium-chloro complexes, and then perform AIMD simulations to explore
more precisely the dynamic nature of the anode/electrolyte salt
interface. Envisioning the overall Mg deposition process, for example,
beginning with a THF-solvated (MgCl)$^+$ complex in the bulk electrolyte
and ending as metallic Mg incorporated in the anode, the deposited Mg
must in some sequence shed its sheath of THF molecules. In
Figure~\ref{fig:binding}, the interfacial structure and corresponding
adsorption energies of several charged magnesium-chloro (MgCl)$^+$
complexes (solvated by 5 through 0 THF molecules and in different
orientations) adsorbed on a Mg(0001) surface are shown. Throughout the
manuscript we adopt an abbreviated nomenclature to discuss
magnesium-chloro (MgCl)$^+$ complexes adsorbed at the Mg surface, namely
the first number indicates the number of THF ligands followed by the
orientation (or site) adopted by the adsorbing (MgCl)$^+$ salt. For
example, 4-SIDE means that (MgCl)$^+$ is coordinated by 4-THF molecules
and lays horizontally on the Mg surface as shown in the snapshots of
Figure~\ref{fig:binding}. Two different liquid references were used to
compute the $\Delta$Es of adsorbed magnesium-chloro salts at the Mg
surface: \emph{i}) bulk (MgCl)$^+$-3THF reference,\cite{Wan2014} and
\emph{ii}) with a bulk (MgCl)$^+$-5THF,\cite{Pour2011,Benzmayza2013} as
indicated respectively by the upper and lower bounds of the bars in
Figure~\ref{fig:binding}. The difference in absorption energies from these
two reference states is small, and can be attributed to the slightly
more stable (MgCl)$^{+}$-3THF reference in the reference calculation.  

Upon first inspection, the magnitude of the energies for salt adsorption
(in Figure~\ref{fig:binding}) compared to THF adsorptions (values shown in
Figure~\ref{fig:adsorption}b and text above) confirms that the THF solvent
weakly interacts with the Mg metal surface while magnesium-chloro salt
complexes readily adsorb, in good agreement with experimental
observations by Benzmayza \emph{et al.}\cite{Benzmayza2013} and the
reversible plating observed in this
system.\cite{Aurbach2002,Pour2011,Doe2014,Barile2014,Yoo2013}  In all
cases, Cl aligns toward the Mg surface, with adsorption on the HLW site
preferred over the TOP site as shown in Figure~\ref{fig:adsorption} (also
see Figure~S2 in Supplementary Information), because in comparison to TOP,
HLW and SIDE oriented adsorption (shown in Figure~\ref{fig:binding})
maximizes the interaction of the Cl ion with the Mg surface sites as
well as that of some dangling C and H atoms from the THF molecules with
the Mg surface. Of all the charged magnesium-chloro complexes,
(MgCl)$^+$ coordinated by 5 THF in the hollow site exhibits the
strongest adsorption energy (-84.7 $\pm$~10.5 kJ~mol$^{-1}$), and in
general the strength of the adsorption reduces with sequential THF
removal, as seen in the trend of adsorption energies of
Figure~\ref{fig:binding}. Also, as THF is successively removed from the
salt complex, the most stable orientation relative to the surface shifts
from HLW to SIDE. Both trends can be understood by considering the role
of the Mg surface in stabilizing known preferred Mg$^{2+}$ coordination
geometries.\cite{Benzmayza2013} For 5-HLW, Mg$^{2+}$ can remain stably
in octahedral coordination as seen in Figure~\ref{fig:binding}, but as
more THF are removed; Mg$^{2+}$ is increasingly exposed to unstable
coordinations. For example, 4-HLW has a more positive adsorption energy
in comparison. Adopting the SIDE orientation allows Mg$^{2+}$ to be
additionally coordinated by the Mg surface, which can stabilize the salt
complex when there are too few THFs to ``ideally'' solvate Mg$^{2+}$ as
seen in Figure~\ref{fig:binding}, where the SIDE configuration becomes the
most stable for 3 THFs or fewer. As even more THF molecules are removed,
the adsorption of the salt complexes is continually weakened, but a
notable exception is from 3-SIDE to 2-SIDE (as seen in
Figure~\ref{fig:binding}) where there is an increase in adsorption
strength likely due to less steric hindrance between THF molecules which
can be relieved by removing the most weakly bound THF. The adsorption of
(MgCl)$^+$-1THF,1-SIDE, and bare (MgCl)$^+$, 0-SIDE on the Mg(0001)
surface were found to be very unfavorable (+96.8 and +31.1 kJ~mol$^{-1}$
respectively), indicating that these species are likely too unstable to
play a significant role in the reversible Mg deposition process. 

\begin{figure*}[t]
\includegraphics[scale=0.5]{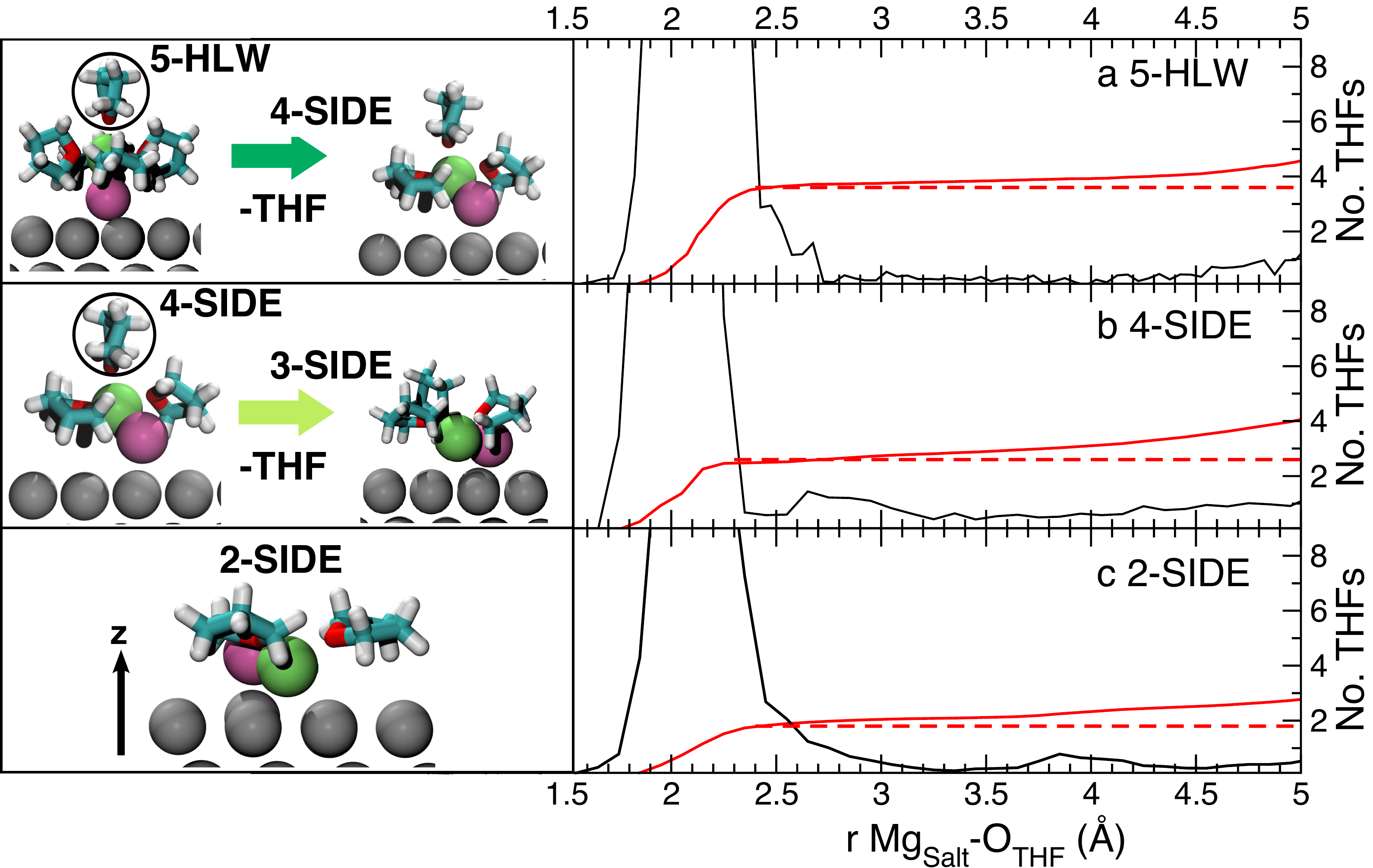}
\caption{\label{fig:RDFsalt} (Color online) PDFs g(r) and coordination
number of Mg atoms in the (MgCl)$^+$ salt and THF oxygen atoms for
models (a) 5-HLW, (b) 4-SIDE and (c) 2-SIDE (see
Figure~\ref{fig:binding}). Insets show the structural evolution of the
salts at the surface. Red dashed lines as guide for the eye. Black
circles indicate THF removal. Mg-Cl PDFs in Figure~S4 of the Supplementary
Information. }
\end{figure*}

As mentioned earlier, the (MgCl)$^+$ monomer surrounded by a solvent
coordination shell of 4 or 5 THF molecules adsorbs strongly to the Mg
surface (as seen in Figure~\ref{fig:binding}), but these complexes also
preserve the integrity of the weak Mg-Cl bond. In addition to electron
transfer, completion of the Mg deposition process also requires the
eventual dissociation of Mg from Cl$^-$ at the Mg surface, which
necessitates breaking the solvation shell. The differences in adsorption
energies shown in Figure~\ref{fig:binding}, can also be considered as the
energy to (de)solvate the (MgCl)$^+$ complex by sequential (removal)
addition of THF molecules, and initial desolvation appears facile with
only a cumulative $\sim$ 46.2 -- 61.0  kJ~mol$^{-1}$ required to remove
3 THFs from the stably adsorbed (MgCl)$^+$-5THF complex.  To investigate
this observation further under more realistic conditions comparable to
experiment, we perform room temperature AIMD simulations with different
(MgCl)$^+$ complexes adsorbed on the Mg surface in a liquid THF
reservoir. Figure~\ref{fig:RDFsalt} shows the pair distribution function
between the Mg atom of the (Mg-Cl)$^+$ salt and the oxygen atoms of the
THF ligands (black lines) and the respective coordination numbers (red
lines) for three different salt adsorption configurations, 5-HLW, 4-SIDE
and 2-SIDE.  In Figure~\ref{fig:RDFsalt} the average Mg-O(THF) distance is
centered around 1.9 -- 2.2 \AA, falling in the typical Mg-O bond length
range (2.1 -- 2.2 \AA). The coordination number measured at the center
of the g(r) peaks (see red dashed lines in Figure~\ref{fig:RDFsalt} as a
guide for the eye) is 4, 3, and 2 for 5-HLW, 4-SIDE and 2-SIDE,
respectively.  

Surprisingly, in some cases the number of THFs in the first coordination
shell of the (MgCl)$^+$ salt spontaneously decreases by one THF unit
after few picoseconds of equilibration time in the AIMD simulation.  For
example, when beginning the AIMD simulation with either 5-HLW or 4-HLW
absorbed on the Mg surface, we observe as the final result 4-HLW and
3-SIDE, respectively, as seen in the dashed red lines in
Figure~\ref{fig:RDFsalt}a and Figure~\ref{fig:RDFsalt}b. Likely, this is due
to additional THF-THF interactions between the ligands of the adsorbed
complex and bulk liquid in the AIMD simulation (unaccounted for in the
static adsorption calculations which are performed in vacuum) and the
effect of entropy, which stabilizes THF in the liquid compared to the
adsorbed state. This evidence further supports Mg ion desolvation as a
facile process. As seen in Figure~\ref{fig:RDFsalt}, the trend of
spontaneously shedding THF stops in the 2-SIDE configuration, as the
significant increase in desolvation energy required to further remove
THF (as seen in Figure~\ref{fig:binding}) cannot be offset by the finite
temperature and additional solvent interactions.  

The large positive energy increases when removing additional THF i.e.\
from 2-SIDE to 1-SIDE to 0-SIDE, suggests that, the electron transfer
likely occurs before, when there are still some surrounding THF
molecules (likely 2), which further underscores the necessity of
considering the stabilizing coordination effect of the solvent ion with
an explicit 1$^{st}$ coordination shell of THF.  From close inspection
of the relaxed geometries of low and zero-coordinated THF complexes
(such as adsorbed 1-SIDE and 0-SIDE, see Figure~\ref{fig:binding}), we
observe incipient dissociation of the Mg-Cl bond (this also occurs for
some of the MgCl$_2$ configurations (see Figure~S5 in the Supplementary
Information), which further suggests that magnesium-chloro complexes
with few THF molecules are the relevant species involved in charge
transfer. Fortunately, the SIDE orientation is most energetically
favorable at lower THF coordinations (as shown in
Figure~\ref{fig:binding}), which allows for the depositing Mg$^{2+}$ to
become better exposed to the anode surface, which should better
facilitate ensuing charge transfer.

Finally, we also consider the adsorption on the Mg-surface of the
charge-neutral MgCl$_2$ species that may precipitate at the anode
interface, for example as undissolved/unreacted MgCl$_2$ during the
preparation and possibly conditioning\cite{Barile2014} of the
magnesium-chloro electrolyte. Previous AIMD simulations by Wan \emph{et
al.}\cite{Wan2014} have shown that MgCl$_2$ in bulk THF solvent remains
coordinated by 2 THF molecules (MgCl$_2$-2THF), which we adopt as the
liquid reference to compute the adsorption energies.  In general, the
$\Delta$Es are always found to be very positive for all the models
considered in this study i.e.\ bare MgCl$_2$ (174.8 kJ~mol$^{-1}$),
MgCl$_2$-2THF (25.0 kJ~mol$^{-1}$), and MgCl$_2$-3THF (28.9
kJ~mol$^{-1}$), indicating that MgCl$_2$ adsorption is not competitive
with the (MgCl)$^+$ salts at equilibrium (see Figure~S6 in Supplementary
Information). 


\section{Discussion}
\label{sec:discussion}
To gain practical insights into the design of electrolytes capable of
supporting reversible Mg deposition in Mg-ion batteries, we performed a
comprehensive first-principles calculations based adsorption study of
the relevant salt and solvent species at the interface of a Mg metal
surface in contact with a solution of Mg$^{2+}$ and Cl$^-$ dissolved in
THF solvent using static adsorption calculations and \emph{ab initio}
molecular dynamics simulations. Although our model relies on a few but
important assumptions, namely a Mg metal anode free of defects,
impurities, our results reveal important features of the system
consistent with reversible Mg deposition and other experimental
observations. 

From static adsorption calculations of the anode/solvent system, THF
exhibits weak interaction with Mg metal, which should leave the surface
unpassivated and therefore accessible to salt adsorption. Furthermore,
the adsorption calculations of (MgCl)$^+$ complexes coordinated by
several THF molecules indicate that solvated salt complexes rather than
solvent species are strongly preferred at the surface, which is
confirmed unambiguously in the AIMD simulations. Linear poly-THF
molecules, however, reported to be the product of AlCl$_3$ catalyzed THF
polymerization,\cite{Barile2014} might show larger interaction with the anode
surface in comparison to THF and linear ethers.

In this investigation we have clarified that oxygen impurities (in
form of MgO) do not affect the adsorption properties of the solvent, but as demonstrated by previous studies{\cite{Gofer2003,Barile2014}} traces of other impurities such as Cl and Al can potentially react with salt species and the anode participating to the passivation of the electrode.

Considering the Mg deposition process, once a (MgCl)$^+$ complex
solvated by some (4~or~5) THF is adsorbed at the Mg surface, breaking
the THF solvation shell to facilitate electron transfer is shown to be
initially facile, requiring only $\sim$ 61 -- 46 kJ~mol$^{-1}$ to remove
3 THF ligands from 5-HLW, for example. Simultaneously, the adsorbed
complex adopts a conformation that brings the adsorbed Mg$^{2+}$ ion
nearer to the Mg surface, rotating from HLW to SIDE orientation, which
should benefit electron transfer and eventual Mg incorporation (by
maximizing the orbitals overlap between Mg$^{2+}$ ions and Mg$^0$ atoms
of the electrode). The energy to remove additional solvent molecules
abruptly increases once there are only two remaining THF molecules, and
(MgCl)$^+$ complexes solvated by 1 and 0 THF molecules do not even bind
to the Mg surface, which strongly suggests that (MgCl)$^+$ coordinated
by 2 THF molecules is the relevant complex involved in charge transfer,
an important finding requiring further experimental verification.  

The electrochemical activity of Cl as an additive to facilitate the
electron transfer during multi-valent metal deposition is well known for
Cu deposition, demonstrated from both a theoretical and experimental
standpoint by Nagy \emph{et al}.\cite{Nagy1995}. The presence of Cl$^-$
accelerates the first electron transfer from Cu$^{2+}$ to Cu$^+$ in the
multistep reduction of copper, and the overlap of Cl$^-$ and Cu$^{2+}$
orbitals eases the electron transfer mechanism by isolating the plating
Cu$^{2+}$ ions from competitive solvent interaction. In that model, it
is crucial that the Cl-Cu$^{2+}$ complex pre-adsorb at the surface, and
indications of the same behavior in this system (preferential adsorption
of (MgCl)$^+$ complexes) are overall in agreement with experimental
observations of reversible Mg deposition.

Although we did not explicitly consider the effect of an applied
electrochemical potential to the system, the adsorption calculations in
this work also shed some light on deposition events that occur post
charge transfer. It is well established that atomic and ionic Cl readily
adsorb in the hollow sites of Mg metal surfaces with highly exothermic
adsorption energies (-241.8 kJ~mol$^{-1}$ calculated in this work and in
agreement with previous work \cite{Cheng2014}), and the observed
stability of isolated Cl species in our calculations suggests that
lingering Cl adsorption on Mg surface after charge transfer may affect
deposition kinetics. Indeed, recent SEM, energy-dispersive X-ray
spectroscopy and XPS measurements reveal traces of Cl (in addition to
Al) at the Mg surface.\cite{Barile2014,Gofer2003} In order to keep
depositing pure Mg metal at the anode, residual Cl must be continually
removed which can contribute to the plating overpotential, and we
speculate that the addition of AlCl$_3$ species to the electrolyte
solution may improve deposition kinetics by ``scrubbing'' the Mg anode
free of Cl.\cite{Doe2014,Barile2014} Specifically, added AlCl$_3$ can
act as a shuttle for Cl$^-$ from the anode to the cathode during Mg
plating by shifting the AlCl$_3$~+ Cl$^-\rightarrow$ AlCl$_{4}^{-}$
reaction equilibrium, as AlCl$_3$ reacts with accumulated Cl$^-$ at the
anode surface, the AlC$_{4}^{-}$ product is driven toward the cathode.
Continual removal of chloride at the anode interface is a source of
overpotential upon deposition, but in the stripping process, the
additional availability of Cl$^-$ should promote the process. During the
dissolution process, the applied potential drives the charged (MgCl)$^+$
complexes to the cathode, which consequently increases the local
concentration of the neutral MgCl$_2$ complexes at the anode, and as we
observe, these spontaneously dissociate into (MgCl)$^+$ and Cl$^-$, thus
creating more carriers to drive the reaction. Overall, this creates a
source of asymmetry between the deposition and stripping process.


\section{Conclusions}
\label{sec:conclusions}
In conclusion, we have set to resolve the atomistic mechanisms that
allow for reversible Mg electrodeposition in non-aqueous electrolytes by
performing a comprehensive first-principles adsorption study at the
interface of a Mg metal anode in contact with an electrolyte solution
consisting of Mg$^{2+}$ and Cl$^-$ dissolved in THF (tetrahydrofuran)
solvent. From an analysis of both solvent and salt adsorption from
static first-principles and \emph{ab initio} molecular dynamics calculations,
we have gained a number of insights as to why this narrow class of
electrolytes (magnesium-chloro salts dissolved in ethereal solvents) can
function in practice: first, neither cyclic ether solvent molecules nor
neutral salt complexes (MgCl$_2$ based complexes) exhibit
surface-passivating behavior (although some polymeric THF shows more
favorable surface interaction in comparison to cyclic THF); second, the
strongest adsorbing species' at the Mg interface are charged (MgCl)$^+$
complexes which are also the active species involved in charge-transfer;
third, the energy to (de)solvate (MgCl)$^+$ is minimal; and finally, the
stable orientations of de-solvated (MgCl)$^+$ complexes are also
favorable for charge transfer. In future work, we plan to directly
investigate the effect of an applied potential on the interfacial
properties considered in this study with the goal of directly
characterizing the electron-transfer mechanism upon Mg deposition and
stripping.

\acknowledgements
This work was fully supported as part of the Joint Center for Energy
Storage Research (JCESR), an Energy Innovation Hub funded by the U.S.
Department of Energy, Office of Science, and Basic Energy Sciences. This
study was supported by Subcontract 3F-31144. We also thank the National
Energy Research Scientific Computing Center (NERSC) for providing
computing resources. This research used resources of the Argonne
Leadership Computing Facility, which is a DOE Office of Science User
Facility supported under Contract DE-AC02-06CH11357, project
1309-REBatteries Revealing the Reversible Electro-deposition Mechanism
in Multivalent-Ion Batteries. PC is grateful to Dr.\ E.\ Torres at MIT
for insight in the use of the DL\_POLY code. PC is grateful to W.\
Richards at MIT, and Dr.\ N.\ Hahn at Sandia National Lab for  fruitful
suggestions and discussions.	


\bibliography{biblio}

\end{document}